\documentclass[entropy,article,accept,pdftex,moreauthors]{Definitions/mdpi}

\firstpage{1}
\pubvolume{27}
\issuenum{6}
\articlenumber{636}
\pubyear{2025}
\copyrightyear{2025}
\externaleditor{Changshui Yu} 
\datereceived{25 May 2025 }
\daterevised{7 June 2025 } 
\dateaccepted{12 June 2025}
\datepublished{14 June 2025}
\hreflink{https://doi.org/ 10.3390/e27060636} 
\pdfoutput=1 

\usepackage{braket}
\usepackage{siunitx}
\usepackage[english]{babel}
\usepackage{csquotes}
\usepackage{dsfont}

\Title{Measurement-Induced Dynamical Quantum Thermalization}

\TitleCitation{Measurement-Induced Dynamical Quantum Thermalization}

\Author{Marvin Lenk $^{1}$\href{https://orcid.org/0009-0000-9844-2103}{\orcidicon}, Sayak Biswas $^{2}$\href{https://orcid.org/0009-0002-7880-2364}{\orcidicon}, {Anna Posazhennikova} 
$^{3}$\href{https://orcid.org/0000-0002-7757-5086}{\orcidicon} and Johann Kroha $^{1,4,}$*\href{https://orcid.org/0000-0002-3340-9166}{\orcidicon}}

\AuthorNames{Marvin Lenk, Sayak Biswas, Anna Posazhennikova and Johann Kroha}

\AuthorCitation{Lenk, M.; Biswas, S.; Posazhennikova, A.; Kroha, J.}

\address{%
$^{1}$ \quad Physikalisches Institut,
Universit\"at Bonn, Nussallee 12, 53115 Bonn, Germany; {mlenk@uni-bonn.de}\\
$^{2}$ \quad Physics Department, The Ohio State University, Columbus, OH 43210, USA; {biswas.124@osu.edu}\\
$^{3}$ \quad Institut für Physik, Universität Greifswald, 17487 Greifswald, Germany; {anna.posazhennikova@uni-greifswald.de}\\
$^{4}$ \quad School of Physics and Astronomy, University of St. Andrews, North Haugh, St. Andrews KY16 9SS, UK
}

\corres{Correspondence: kroha@physik.uni-bonn.de}

\abstract{One of the fundamental problems of quantum statistical physics is how an ideally isolated quantum system can ever reach thermal equilibrium behavior despite the unitary time evolution of quantum-mechanical systems. Here, we study, via explicit time evolution for the generic model system of an interacting, trapped Bose gas with discrete single-particle levels, how the measurement of one or more observables subdivides the system into observed and non-observed Hilbert subspaces and the tracing over the non-measured quantum numbers defines an effective, thermodynamic bath, induces the entanglement of the observed Hilbert subspace with the bath, and leads to a bi-exponential approach of the entanglement entropy and of the measured observables to thermal equilibrium behavior as a function of time. We find this to be more generally fulfilled than in the scenario of the eigenstate thermalization hypothesis (ETH), namely for both local particle occupation numbers and non-local density correlation functions, and independent of the specific initial quantum state of the time evolution.}


\newcommand*{\dphant}{{\ensuremath{\vphantom{\dagger}}}}

\begin{document}
%


\section{Introduction}

Thermodynamics, as a phenomenological theory, has proven to be most useful since its inception. It was constructed to describe macroscopic systems in contact with each other, where concepts like heat exchange, entropy, and temperature are defined in an operative way. In this context, thermodynamic equilibrium emerges empirically as the state that is reached after a sufficiently long time when a system is left to itself by maximizing its entropy. Such an equilibrium state can be realized by heat exchange with a thermal heat bath at a given temperature or in an isolated, classical system at constant energy by multiple scattering of the microscopic entities comprising the macroscopic system.
In an isolated quantum system, however, the problem of thermalization is non-trivial because quantum-mechanical time evolution of isolated or Hamiltonian systems is unitary: A statistically pure quantum state remains pure, and, therefore, its entropy is equal to zero for all times. More generally, in an isolated quantum system, the total entropy is constant in time and cannot be maximized. In sharp contrast, experiments on trapped, ultracold atomic gases, which are nearly ideally isolated from the environment, do show exponential relaxation to thermal behavior. This constitutes the problem of quantum thermalization.

One of the most prominent scenarios for understanding the thermal behavior of isolated quantum systems is the eigenstate thermalization hypothesis (ETH), put forward independently by Deutsch and Srednicki \cite{deutsch_quantum_1991,srednicki_chaos_1994}. It asserts that in an isolated many-body system, the long-time average of the quantum-mechanical expectation value
{$\bra{\Psi(t)}\hat{A}\ket{\Psi(t)}$} of an observable quantity $\hat{A}$ with respect to a statistically pure state $\ket{\Psi(t)}$ is indistinguishable from its expectation value with respect to an energy eigenstate at a typical energy and, thus, indistinguishable from the microcanonical thermal average. The ETH became a focus of quantum research after its numerical verification in hard-core boson gases in two-dimensional lattices \cite{rigol_ETH_2008,rigol_ETH_2016}, although it was found not to be valid in other systems \cite{rigol_no-ETH_2009,pozsgay_2014}. The applicability of the ETH relies on a number of restrictive preconditions. Most notably, the energy distribution of the Hamiltonian eigenstates comprising the pure quantum state $\ket{\Psi(t)}$ must be narrow compared to its mean energy, in order to fulfill the microcanonical condition. In addition, the off-diagonal elements of $\bra{\Psi(t)}\hat{A}\ket{\Psi(t)}$ in the energy eigenbasis must effectively vanish. This, however, occurs generally only on the long-time average by destructive interference, which is slow due to the narrow frequency spectrum, in contrast to the experimentally observed, fast thermalization dynamics of ergodic systems.  See Ref.~\cite{Posazhennikova2018} for a detailed discussion and the relation to ergodicity. Typically, the ETH can be validated only for local observables (correlation functions) \cite{Eccles2025}.

In the present work, we explore the thermalization mechanism of measurement-induced, dynamical (heat) bath generation (DBG), put forward in Refs.~\cite{Posazhennikova2016,Posazhennikova2018}.
It is based on the fact that in a sufficiently complex system (sufficiently complex means that the Hilbert space dimension is large and the Hilbert space does not factorize into disjoint sectors by the system dynamics),
it is not possible to completely determine the state vector, i.e., to measure all quantum numbers of the system by any realistic experiment. Thus, the measurement subdivides the Hilbert space into an observed and a non-observed sector, such that the latter acts as a canonical reservoir for the observed subsystem. As a consequence, the entanglement entropy of the reduced state vector will be maximized even though the total system remains in a pure state. This concept was introduced by Goldstein {et al.} by the term \textit{{canonical typicality}}
\cite{goldstein_canonical_2006}, building on earlier ideas of intrinsic decoherence in a quantum system by Zurek \cite{zurek_pointer_1981, zurek_environment-induced_1982, zurek_decoherence_2003} and by Milburn  {et al.} \cite{milburn_intrinsic_1991}, and further developed by \linebreak  Yukalov \cite{Yukalov2012_thermalization,Yukalov2012_decoherence}. Rigorous results on thermalization due to canonical typicality and its relation to the ETH in discrete, translationally invariant systems were proven in Ref.~\cite{Mueller2015}. The work of Posazhennikova {et al.}~\cite{Posazhennikova2016,Posazhennikova2018} extended this concept of typicality from stationary, thermal states to the time-dependent approach to equilibrium, then dubbed as DBG, albeit the calculations were performed for overall grand-canonical systems, i.e., effectively open \linebreak  quantum systems.

We examine the validity of the predictions of the DBG by numerically exact time evolution of closed quantum many-body systems, where the total energy and particle number are conserved, and monitoring the temporal approach to thermal equilibrium behavior. As a generic, non-integrable system, we consider a Bose gas of up to $N=25$ particles, confined in a trap comprising five single-particle levels and with an arbitrarily strong two-particle repulsion $U$. For this system, the natural observables are the (time-dependent) occupation numbers of the individual single-particle levels as well as the occupation-number correlation functions between different levels.
We reach exponentially long evolution times by iteratively exponentiating the time-evolution operator. The exponentiation procedure requires computing the time-evolution operator for the entire Hilbert space, not only the components comprising the state vector at hand, as in the adaptive, time-dependent density-matrix renormalization group (t-DMRG) \cite{Kollath2007}. Although this does limit the system size, it turns out that even our few-particle systems of $N\leq 25$ are complex enough to show clear thermal behavior after a sufficiently long time. This indicates, in particular, that the thermodynamic limit of infinite system size is controlled by the Hilbert space dimension rather than the number of particles. We further find that the entanglement entropy of the observed subsystem with the non-observed one approaches a maximum with a bi-exponential time dependence, independent of the energy distribution of the initial state. We show that this entropy maximum represents a thermal state by fitting the fluctuation--dissipation relation to the single-particle spectra as well as the level--occupation correlation functions. As an important result, we find that not only the local correlation functions thermalize in this sense but also the non-local correlation functions of occupation numbers of different levels.

The paper is organized as follows.
In the following Section~\ref{sec:DBG}, we give a pedagogical description of the partitioning of an arbitrary quantum system by the measurement characterized by the reduced density matrix, whose entanglement entropy may be maximized although the entire system remains in a pure quantum state. 
In Section~\ref{sec:model_method} we introduce, as a generic model for the concrete calculations, a trapped, interacting Bose gas and outline our computational method of long-time evolution. Section~\ref{sec:entropy} contains results for the entanglement entropy of the observed subsystem and analyzes its approach to a maximum as a function of time for initial pure states with a narrow (microcanonical) as well as a wide (canonical) energy distribution. In Sections \ref{sec:loc_occ_numbers} and \ref{sec:nonlocal_correlations}, we present the thermalization dynamics of the boson occupation numbers in specific single-particle levels and the correlation functions between the occupation numbers of different levels, non-local in level space, respectively. We conclude with a discussion and outlook of the general applicability of our results in Section  \ref{sec:conclusion}.
\section{Measurement and Entropy in Isolated Quantum Systems}\label{sec:DBG}
The density matrix of an isolated quantum system with the Hamiltonian $\hat{H}$ reads,
\begin{align}
\hat{\rho}(t) = \sum_{\alpha, \beta} P_{\alpha\beta}   \ket{\psi_{\alpha}(t)} \bra{\psi_{\beta}(t)}\,,
\label{eq:rho_general}
\end{align}
where the sum runs over the orthonormal basis states $\ket{\psi_{\alpha}(t)}$ of the complete many-body system. In the Schr\"odinger picture, their time dependence, $\ket{\psi_{\alpha}(t)}=\hat{U}(t)\ket{\psi_{\alpha}(0)}$, is given by the time-evolution operator $\hat{U}={\rm e}^{-i\hat{H}t}$ ($\hbar =1$ here and in the following). As a consequence, the von Neumann entropy of the entire system,
\begin{align}
S(t) = -{\rm tr}\{\hat{\rho}(t) \ln \hat{\rho}(t)\} = -{\rm tr} \{ \hat{U}(t) \; \hat{\rho} \ln \hat{\rho} \; \hat{U}^{-1}(t)\} =S(t=0)\,,
\end{align}
does not depend on time due to cyclic permutation under the trace. In particular, the entropy of a pure state, whose density matrix has only one non-zero eigenvalue, is zero for all times. The ETH uses the fact that in the energy eigenbasis of the many-body system, $\ket{\psi_{\alpha}(t)}={\rm e}^{-iE_{\alpha}t}\ket{\psi_{\alpha}(0)}$, the off-diagonal elements of $\hat{\rho}(t)$ in Equation~(\ref{eq:rho_general}) oscillate with the frequency difference of the respective eigenstates. As a consequence, the off-diagonal elements of $\hat{\rho}(t)$ average to zero in the long-time integral, while the diagonal elements are constant. In the long-time average, the expectation value of an observable $\hat{A}$ can then be written in the energy eigenbasis as the sum over diagonal elements,
\begin{align}
\langle A\rangle = {\rm tr}\{\hat{\rho}\hat{A}\}=\sum_{\alpha} \bra{\psi_{\alpha}}\hat{\rho}\ket{\psi_{\alpha}}\bra{\psi_{\alpha}}\hat{A}\ket{\psi_{\alpha}}.
\end{align}

With the assumption of ergodicity (equivalence of temporal and ensemble average), this can be seen as a thermal average of the observable $\hat{A}$ (see \cite{deutsch_quantum_1991,srednicki_chaos_1994,Posazhennikova2018} for details). However, the off-diagonal elements decay as $1/t$ only after long-time $t$ integration, and a faster diagonalization of the density matrix is possible only on average over all off-diagonal elements. The dynamical approach to equilibrium from an arbitrary initial state is, therefore, unclear in this scenario.

\begin{figure}[H]
\includegraphics[width=0.48\textwidth]{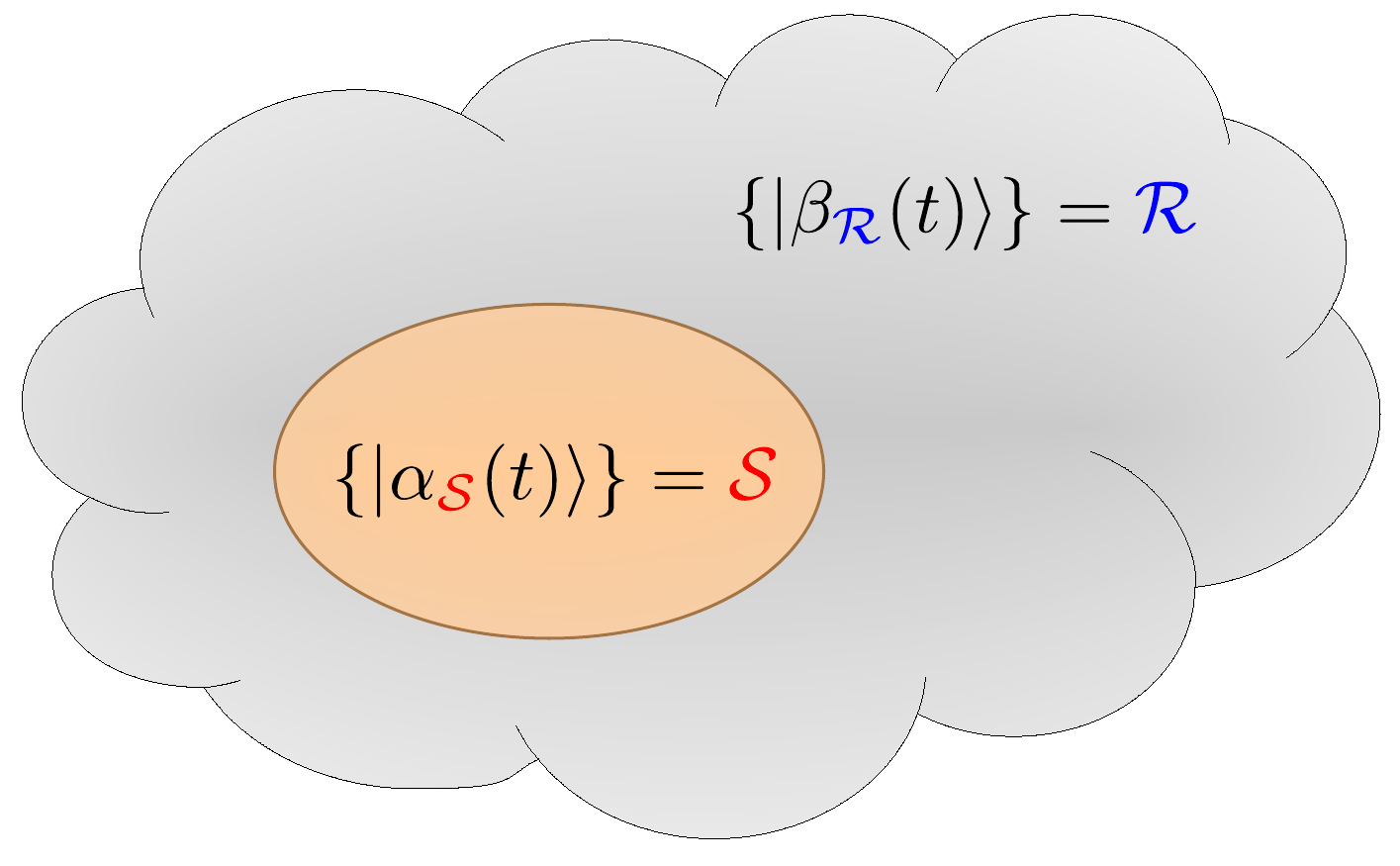}
\caption{{A graphical}
illustration of the measurement-induced partitioning. The cloud represents the full Hilbert space $\mathcal{H}$. The subsystem of interest $\mathcal{S}$ is depicted as an ellipse living inside $\mathcal{H}$. The complement of the ellipse in the cloud is the effective ``reservoir'' subsystem $\mathcal{R}$.}
\label{fig:partitioning}
\end{figure}

Here, we explore the scenario of measurement-induced partitioning. If a quantum system is sufficiently complex, more precisely, the Hilbert space dimension is sufficiently large, it is not possible to completely determine the system's state vector by any given experiment. The measurement of an observable $\hat{A}$ then partitions the complete Hilbert space into the subspace spanned by the quantum numbers determined by the measurement, termed \textit{{system}} $\cal{S}$, and the subspace of not observed quantum numbers which we call \textit{{bath}
} or \textit{{reservoir}} $\cal{R}$, as illustrated in Figure~\ref{fig:partitioning}.

\vspace{-3pt}
The observable $\hat{A}$ can, thus, be decomposed as $\hat{A} = \hat{A}_{\mathcal{S}} \otimes \mathds{1}_{\mathcal{R}}$, where $\hat{A}_{\mathcal{S}}$ acts in the observed subspace only.  Denoting arbitrary orthonormal basis sets of the system and of the reservoir by {$B_{\mathcal{S}}=\{\ket{\alpha_\mathcal{S}}\}$} and $B_{\mathcal{R}}=\{\ket{\beta_\mathcal{R}}\}$, respectively, the complete density \mbox{matrix reads}
\begin{align}
\hat{\rho}(t) &= \sum_{\alpha_{\mathcal{S}},\beta_{\mathcal{R}},\ \alpha'_{\mathcal{S}},\beta'_{\mathcal{R}}} c_{\alpha_{\mathcal{S}}\beta_{\mathcal{R}}} c^*_{\alpha'_{\mathcal{S}}\beta'_{\mathcal{R}}} \ket{\alpha_{\mathcal{S}}\beta_{\mathcal{R}}} \bra{\alpha'_{\mathcal{S}}\beta'_{\mathcal{R}}}\,,
\label{eq:rho_partitioned}
\end{align}
with the direct product states $\ket{\alpha_{\mathcal{S}}\beta_{\mathcal{R}}}= \ket{\alpha_{\mathcal{S}}} \otimes \ket{\beta_{\mathcal{R}}}$ and the sums run over the complete system or reservoir basis, respectively. Since the observable $\hat{A}$ acts only in the observed subspace, $\hat{A} = \hat{A}_{\mathcal{S}} \otimes \mathds{1}_{\mathcal{R}}$, its expectation value, valid for all times $t$, can be written as
\begin{align}
\langle A\rangle (t) = {\rm tr}_{\mathcal{S}}\{\hat{\rho}_{\mathcal{S}}(t) \hat{A} \}\,,
\end{align}
where the reduced system density matrix and its entanglement entropy read, respectively,
\begin{align}
\rho_{\mathcal{S}} (t) &= {\rm tr}_{\mathcal{R}}\{\hat{\rho}(t)\}=\sum_{\tilde{\beta}_{\mathcal{R}}} \sum_{\substack{\alpha_{\mathcal{S}},\beta_{\mathcal{R}}\\\alpha'_{\mathcal{S}},\beta'_{\mathcal{R}}}}
c_{\alpha_{\mathcal{S}}\beta_{\mathcal{R}}} c^*_{\alpha'_{\mathcal{S}}\beta'_{\mathcal{R}}} \bra{\tilde{\beta}_{\mathcal{R}}}\hat{U}(t)\ket{\alpha_{\mathcal{S}} \beta_{\mathcal{R}}} \bra{\alpha'_{\mathcal{S}} \beta'_{\mathcal{R}}} \hat{U}^{-1}(t) \ket{\tilde{\beta}_{\mathcal{R}}},
\label{eq:rho_reduced} \\
S_{\mathcal{S}}(t) &= - {\rm tr}_{\mathcal{S}}\{\hat{\rho}_{\mathcal{S}}(t)  \ln \hat{\rho}_{\mathcal{S}}(t) \} \,.
\label{eq:entanglement_entropy}
\end{align}

Importantly, if the Hamiltonian $H=H_{\mathcal{S}}+H_{\mathcal{R}}+H_{\mathcal{SR}}$ couples system and bath, \linebreak  e.g., by particle exchange induced by $H_{\mathcal{SR}}$, cyclic permutation of the time-evolution operator $\hat{U}(t)$ in Equation~(\ref{eq:rho_reduced}) is not allowed. This is due to the noncommutativity of the action of $\hat{U}(t)$ and the incoherent summation over bath states, $\sum_{\tilde{\beta}_{\mathcal{R}}}$, in Equation~(\ref{eq:rho_reduced}).
Consequently, for an arbitrary initial state, the entanglement entropy will generally be time-dependent. It is the purpose of the present paper to analyze, by numerical evaluations, if and how it may approach a maximum and lead to thermal behavior of physical quantities in the subsystem $\mathcal{S}$. In the following, we will drop the subscript $\mathcal{S}$ on the entanglement entropy, since the time-independent, total entropy will not be further used.
\section{Model and Time-Evolution Method}\label{sec:model_method}
As a concrete model for the numerical time evolution, we analyze a generic, isolated, non-integrable many-body system of $N$ bosons in a trap made up of $M=5$ single-particle~states,
\begin{align}
\hat{H} = \Delta \sum_{i=1}^M (i - 1) \; b_i^\dagger b_i^\dphant + J \sum_{i=1}^M \sum_{j\neq i} b_i^\dagger b_j^{\dphant}
+ U \sum_{i=1}^M b_i^\dagger b_i^\dagger b_i^\dphant b_i^\dphant + U^\prime \sum_{\substack{i,j,l,m\\ \text{\tiny not i=j=l=m}}}^M b_i^\dagger b_j^\dagger b_l^\dphant b_m^\dphant\,,\label{eq:hamiltonian}
\end{align}
where $U$, $U'$ denote the intra-level and inter-level interactions, respectively, and $J$ is the transition amplitude between any pair of levels. $\Delta$ is the single-particle level spacing, assumed to be non-degenerate and equidistant, corresponding to a one-dimensional, harmonic trap. In the present paper, we will show results for the parameter values $\Delta / J = 10$, $U / J = 1$, $U^\prime / J = 0.1$, and a conserved, total particle number $N=25$.  For this parameter set, the average ratio of consecutive level spacings of the many-body eigenenergies is $\langle r \rangle = 0.53$, close to the value for the Gaussian orthogonal ensemble (GOE) of random matrices of $\langle r \rangle =0.5307$ \cite{oganesyan_localization_2007,atas_distribution_2013}, indeed indicating the chaoticity of our many-body system. We also performed calculations for different parameter values, including $N<25$, and the particular choice is not important for the physics described here. Natural observable quantities for our model system will be the time-dependent, local occupation numbers of the single-particle levels, $\langle n_i \rangle (t)$, as well as their non-local correlation functions, $\langle n_i(t) n_j(t+\tau)\rangle $.

To propagate an initial state to the sufficiently long times $t$ necessary to observe the approach of the entanglement entropy to a stationary, maximum value, we use the method of repeated exponentiation of the short-time-evolution operator $\hat{U}_0:=\hat{U}(\delta t)=\exp({-i\hat{H}\delta t})$.  For a short time interval $\delta t$, $\hat{U}_0$ is expanded up to order $k_{\rm max}$,
\begin{align}
\hat{U}(\delta t) = e^{-i \hat{H} \delta t} = \sum_{k=0}^{k_{\rm max}} \frac{(-i \delta t)^k}{k!} \hat{H}^k + \mathcal{O}[(\delta t)^{k_{\rm max}+1}]\,,
\end{align}
where we choose $\delta t$ such that $\delta t \cdot {\rm max} (\hat{H}) \lesssim 10^{-1}$, with ${\rm max}(\hat{H})$ the maximum matrix element of $\hat{H}$ in the entire, large, but finite-dimensional Hilbert space. For $\hat{U}(\delta t)$, this yields a precision of $\mathcal{O}(10^{-k_{\rm max}})$. We usually choose $k_{\rm max}=4$. To reach long evolution times, $\hat{U}_0$ is recursively exponentiated according to
\begin{align}
\hat{U}_1 &= \hat{U}(n\,\ \delta t)=(\hat{U}_0)^n, \nonumber\\
\hat{U}_2 &=\hat{U}(n^2 \delta t) =(\hat{U}_1)^n, \label{eq:t-evolution}\\
\hat{U}_3 &=\hat{U}(n^3 \delta t)=(\hat{U}_2)^n, \qquad {\rm etc.} \nonumber
\end{align}

\textls[-15]{In this way, exponentially long evolution times $\Delta t = n^r\delta t$ are realized at the numerical cost of $(r\cdot n)$ matrix multiplications, growing only linearly with the number of recursions $r$. The precision does not significantly deteriorate for the relevant number of multiplications. A desired, high temporal resolution at long times $\Delta t$ can be achieved by applying lower-order time-evolution operators, after this long time has been reached via high-order evolution operators. We specifically use this approach in the calculation of two-time expectation values. In contrast, single-time expectation values are evaluated on a dense time series utilizing smaller steps with $r \approx 10$ recursions of $n=2$. Note that this procedure is possible because the time-evolution operator does not depend on the system's initial state and can be multiplicatively applied, in contrast to adaptive methods like the adaptive t-DMRG. Conversely, for the recursive multiplication, the evolution operators $U_r$ must be calculated for the entire Hilbert space. For our system of up to $N=25$ bosons in $M=5$ single-particle levels, the Hilbert space dimension is $\approx$24,000, and the matrix multiplications are well feasible}.

\begin{figure}[H]
\includegraphics[width=0.7\linewidth]{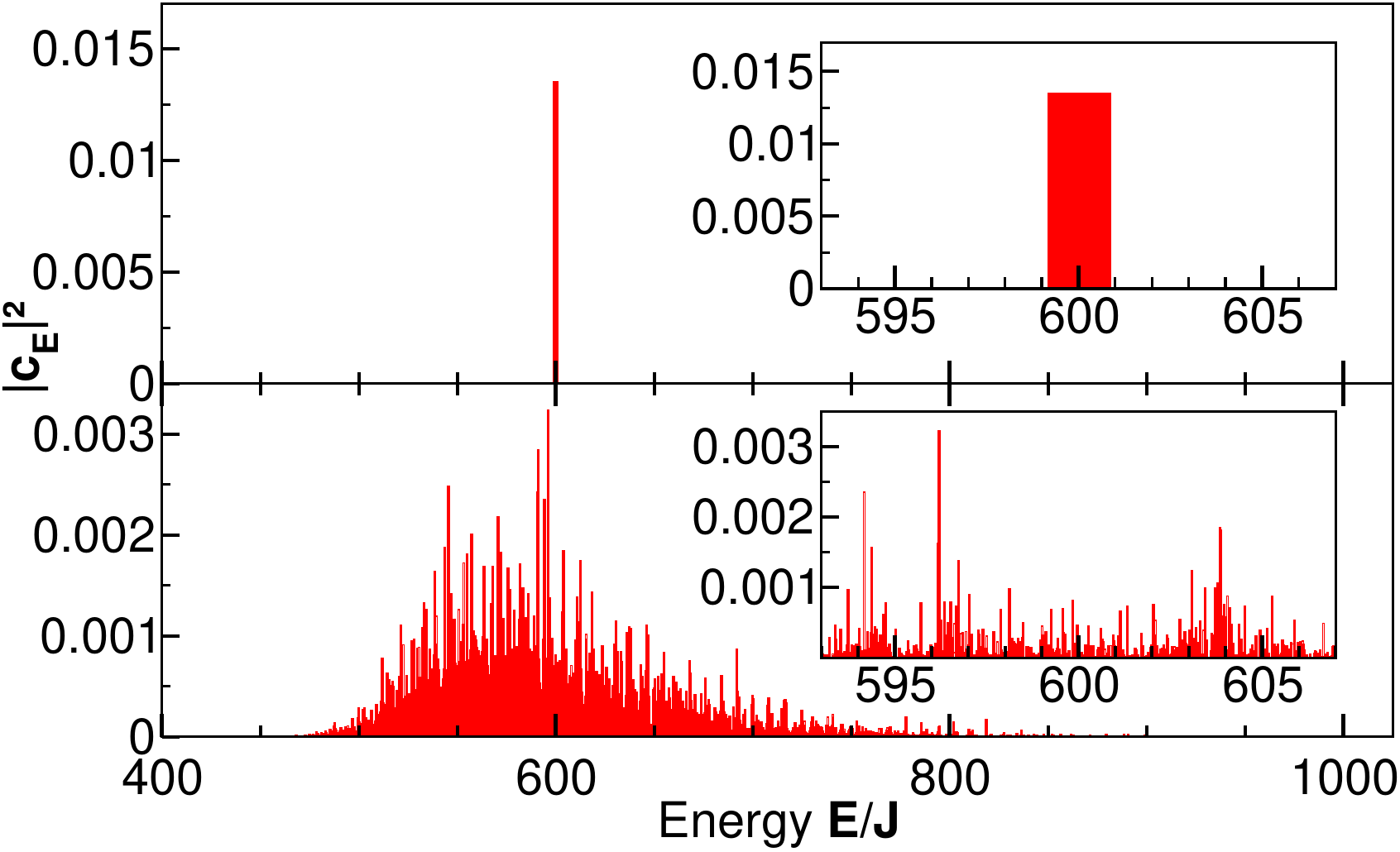}
\caption{Energy spectra of two initial states. The upper panel shows the spectrum of a sharply distributed state compatible with ETH. It is an equal-weighted coherent superposition of the \linebreak  $74$ energy eigenstates between $E/J=599.18$ and $E/J = 600.87$. The lower panel shows the spectrum of the occupation-number eigenstate with all particles in the single-particle ground state, $i=1$. The energy expectation values of both states are equal, $\braket{E}/J = 600$. The insets show small regions around the expectation value.}
\label{fig:spectra}
\end{figure}

An important choice is that of the initial state. In the present study, we consider only pure states of the entire system $\mathcal{S}\otimes\mathcal{R}$, as those states pose the thermalization problem in the most extreme form: the evolution from a zero-entropy state to thermal behavior at long times. We note in passing that, in this case, the entanglement entropy of the system $S_\mathcal{S}$ is identical to the entropy of the rest $\mathcal{R}$~\cite{araki_entropy_1970}.
Among pure states, we investigated two classes: states whose energy spectrum is narrow, $\Delta E \ll \braket{E}$, and states with broad energy spectrum, $\Delta E \lesssim \braket{E}$. The spectrum is defined as the distribution of expansion coefficients of the initial state $\ket{\psi_0}$ into the (discrete) energy eigenbasis via $\ket{\psi_0} = \sum_E \ket{E} \braket{E|\psi_0} = \sum_E c_E \ket{E}$. Narrow-spectrum states are approximate energy eigenstates, i.e., coherent superpositions of many-body eigenstates with nearly equal energy, which may be compatible with the ETH \cite{Posazhennikova2018}. For the broad-spectrum states, we chose occupation-number eigenstates of single-particle levels, which are not Hamiltonian eigenstates due to the transition amplitude $J$ and the inter-level interaction $U^\prime$. Such states do not fulfill the ETH conditions. The spectra of representative narrow-band and broad-band states are shown in Figure~\ref{fig:spectra}.

\section{Entanglement Entropy}\label{sec:entropy}

\textls[-25]{Here, we present the entanglement entropy $S(t)$ for the subsystem $(\mathcal{S}:\, i\geq 3)$ comprising the levels $i\geq 3$ of our system (i.e., the occupation numbers of the levels $i=3,\,4,\,5$ are the measured observables), as computed using the method described in Section~\ref{sec:model_method}. }

\subsection{Broad Energy Spectrum} \label{sec:entropy_wide}
\begin{figure}[H]
\includegraphics[width=0.6\linewidth]{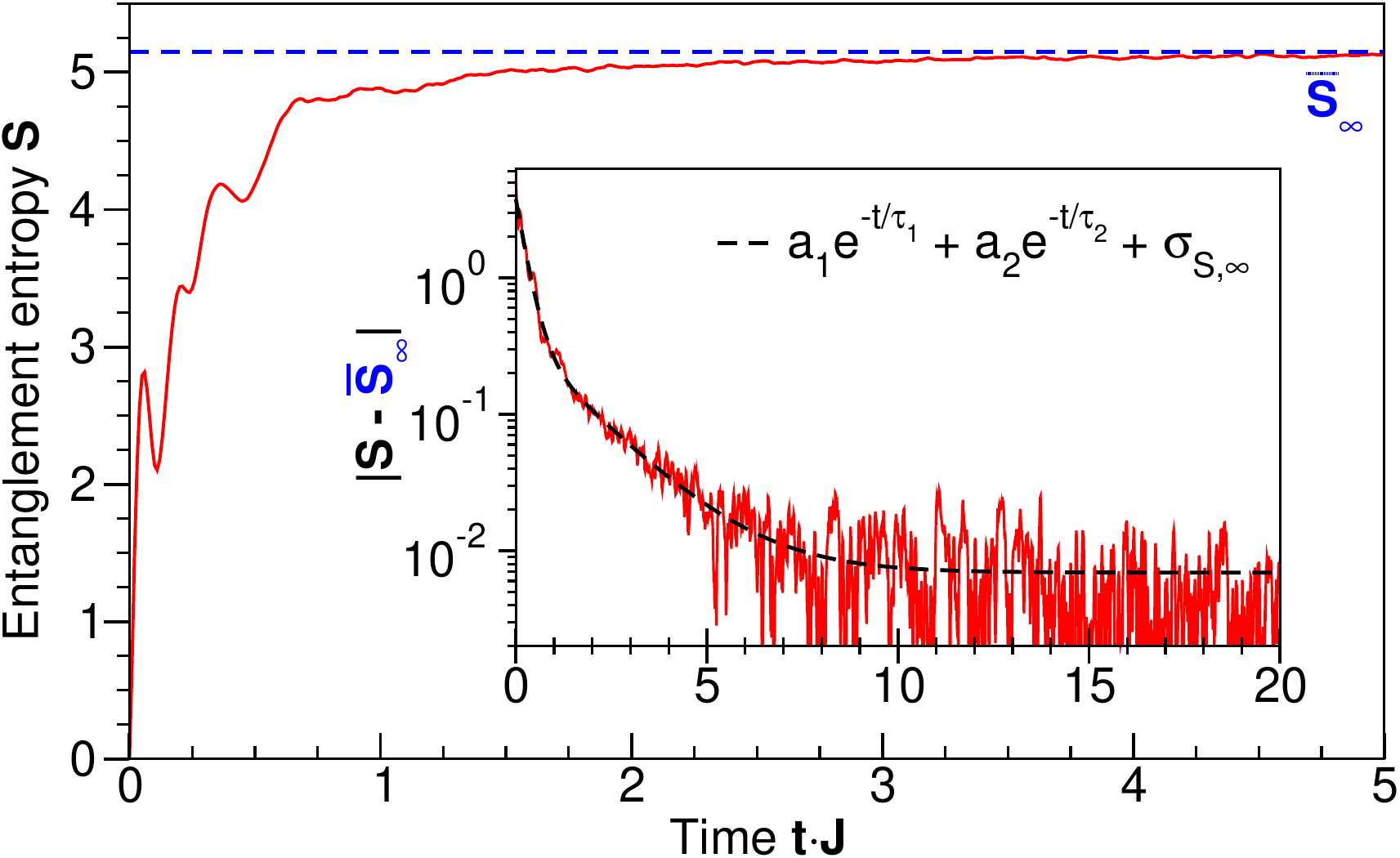}
\caption{{The time}
evolution of the entanglement entropy of the subsystem $\mathcal{S}$ comprising single-particle levels $i \geq 3$. Initial state: occupation-number eigenstate with all particles in the lowest single-particle level, $n_1=N=25$, as shown in Figure~\ref{fig:spectra}, lower panel. The blue dashed line indicates the long-time limit $\bar{S}_{\infty} = 5.15$. In the inset, the modulus of the difference to $\bar{S}_{\infty}$ is shown on a logarithmic scale. The dashed line is a bi-exponential fit, as shown in the legend. The long-time value is the standard deviation calculated directly from the time series as $\sigma_{S,\infty} = 6.9 \cdot 10^{-3}$. The fit parameters are $a_1 = 3.4 \pm 0.2$, $a_2 = 0.35 \pm 0.02$, $\tau_1 J = 0.26 \pm 0.01$, and $\tau_2 J = 1.58 \pm 0.03$.}
\label{fig:entropy}
\end{figure}

Figure~\ref{fig:entropy} shows the time evolution starting from the occupation-number state with $n_1=N=25$, $n_i=0$ for $i>1$, as the broad-spectrum, non-equilibrium initial state (cf.~Figure~\ref{fig:spectra}, bottom panel). Clearly, the entanglement entropy of this initial state vanishes, $S(t=0)=0$. Entanglement is then generated on a fast time scale for times below $t J \approx 1$, followed by a slow, asymptotic approach to a stationary, maximum value attained at $t J \approx 10$ (see inset of Figure~\ref{fig:entropy}). Plotting the deviation from the long-time limit on a logarithmic scale in the inset of Figure~\ref{fig:entropy} reveals two logarithmic slopes, that is, bi-exponential behavior with fitted time constants $\tau_1 J \approx 1/4$ and $\tau_2 J \approx 3/2$ (see also Ref.~\cite{Posazhennikova2016}).

\subsection{Narrow Energy Spectrum}\label{sec:entropy_narrow}
\begin{figure}[H]
\includegraphics[width=0.6\linewidth]{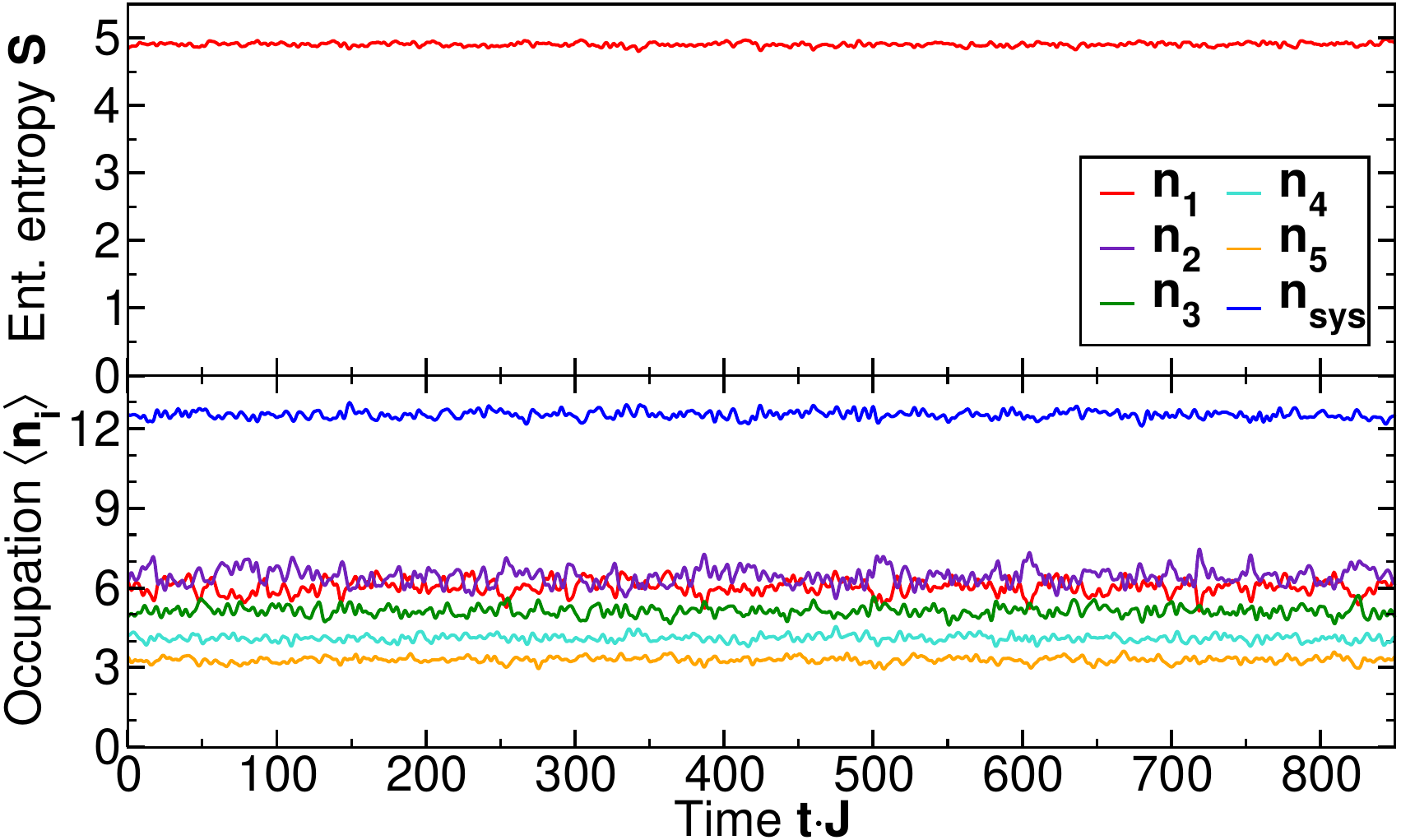}
\caption{{The time} evolution of the entanglement entropy of the subsystem comprising levels $i \geq 3$ and occupation numbers for the microcanonical state, whose energy spectrum is shown in the upper panel of Figure~\ref{fig:spectra}. $n_{\rm sys}$ represents the total particle number in the subsystem $i\geq 3$.}
\label{fig:time_evolution_eth}
\end{figure}

For comparison, we show in Figure~\ref{fig:time_evolution_eth} the time evolution for a microcanonical state as the initial state, i.e., a state with the narrow energy distribution of Figure~\ref{fig:spectra} (upper panel). The initial entanglement entropy for this initial state is already significantly greater than zero, because each energy eigenstate comprising the microcanonical ensemble extends over the entire system, $i=1,\, 2,\, 3,\, 4,\, 5$. The time evolution of the entanglement entropy in the subsystem $(\mathcal{S}:\, i\geq 3)$ is essentially static (upper panel). The same is true for the occupation numbers of the individual single-particle levels $n_i$ (lower panel). This static behavior is expected since the microcanonical state contains essentially only a single frequency $\braket{E}$. The associated time-dependent phase factors cancel in the system density matrix $\hat{\rho}_{\mathcal{S}}$ as well as in the observable expectation values $n_i$. Note that, in this microcanonical state, the distribution of level occupation numbers $P(n_i)$ is not predefined by a temperature but instead fixed by the total energy $\braket{E}$ of the system. A state of this nature is well described by ETH.
At the same time, such a state is essentially static and cannot describe the dynamic approach to a maximum entropy state, since it effectively contains one single frequency only. By contrast, a state with broad energy distribution is not within the scope of the ETH, while the measurement-induced partitioning into subsystem $\mathcal{S}$ and effective reservoir $\mathcal{R}$ covers both regimes, including the dynamical maximization of the entanglement entropy.

\begin{figure}[H]
\includegraphics[width=0.58\linewidth]{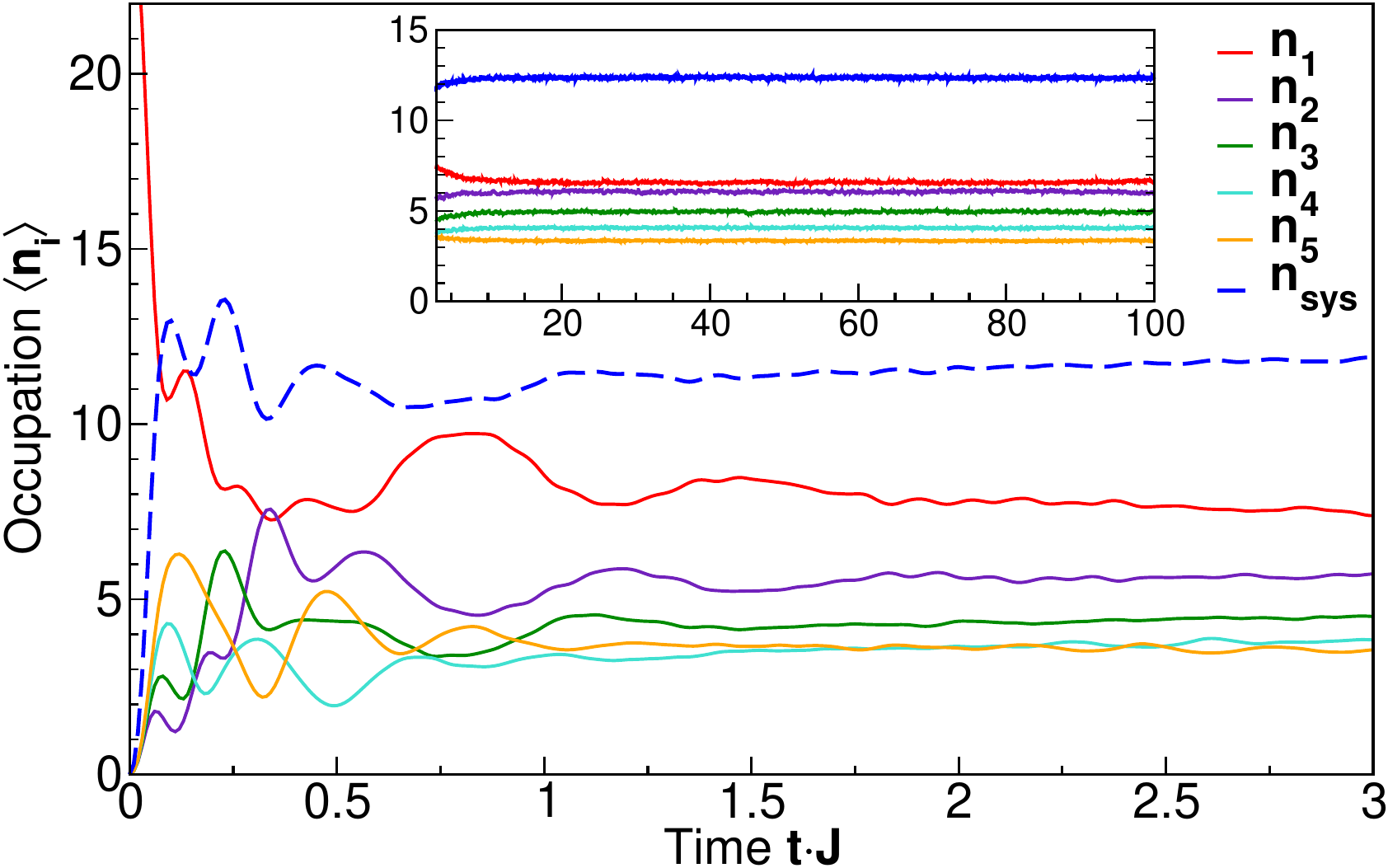}
\caption{The time-dependent occupation numbers of the system comprising three levels \mbox{$n_{\mathrm{sys}} = n_3 + n_4 + n_5$} (blue, dashed line) and individual levels $n_i$.~The inset shows the long-time behavior. Initial state: occupation-number eigenstate shown in Figure~\ref{fig:spectra}, lower panel.}
\label{fig:occupations}
\end{figure}

\section{Thermalization Dynamics of Local Occupation Numbers and Spectra}\label{sec:loc_occ_numbers}

\subsection{Thermalization Dynamics of Local Occupation Numbers}\label{sec:thermalization_occnumbers}
The approach to a steady state and maximum of the entanglement entropy is a necessary but not sufficient condition for thermalization. In this section, we study how physical observables indeed attain a thermal distribution for sufficiently long times. In our trapped Bose gas, the natural, local observables (i.e., on an individual single-particle level) are the occupation numbers $n_i$ and their energy-resolved spectra. Note that measuring or calculating individual occupation numbers defines a different, single-level subsystem for each measurement, while the other levels are traced out.

\begin{figure}[H]
\includegraphics[width=0.58\linewidth]{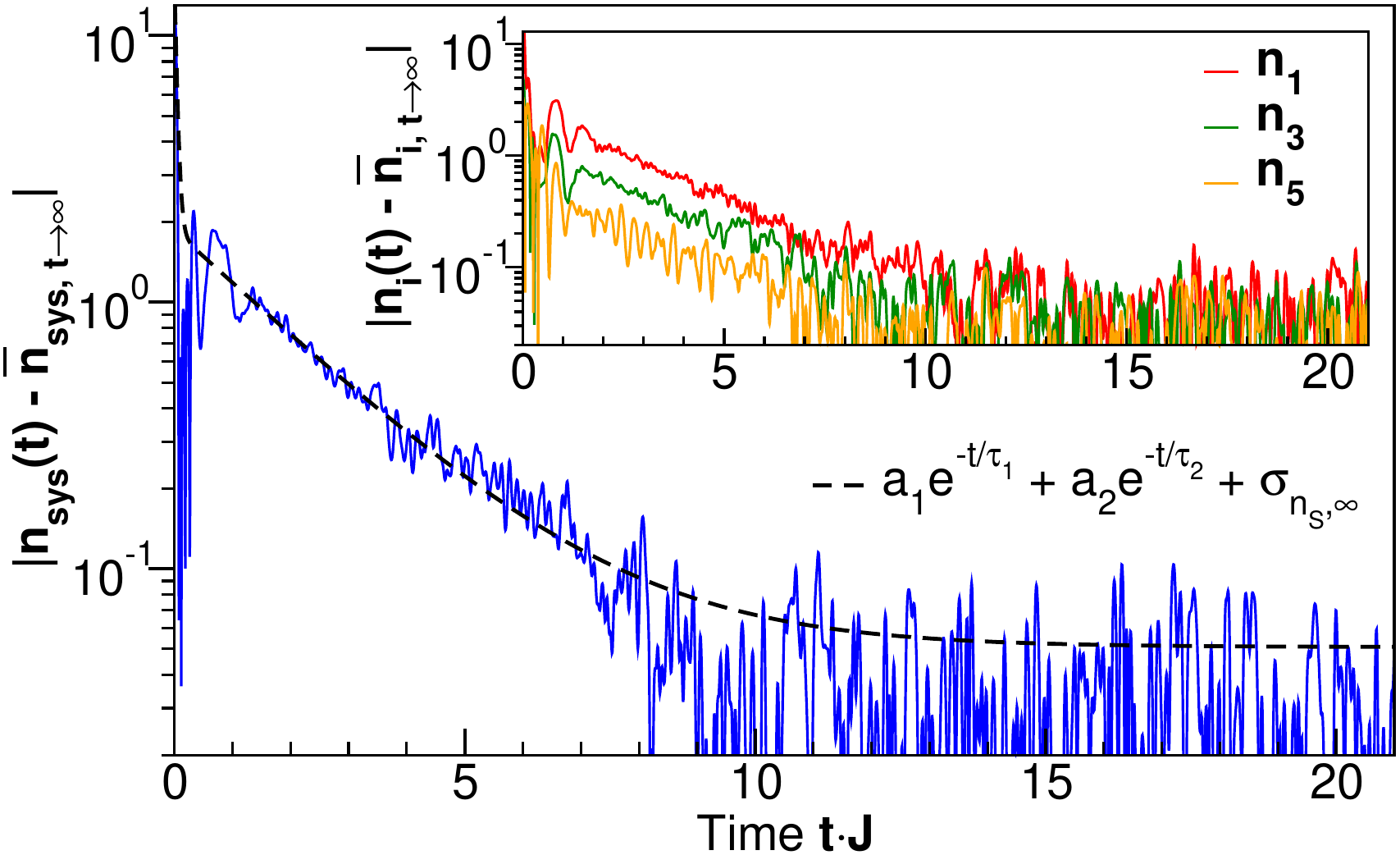}
\caption{{The time}-dependent modulus \textls[-15]{of the deviation of the time-dependent system occupation number $n_{\mathrm{sys}} = n_3 + n_4 + n_5$ from its long-time limit for the occupation-number eigenstate shown in Figure~\ref{fig:spectra}, lower panel, on a logarithmic scale. The inset shows the same quantity for three different single-site occupations. The dashed line in the main panel is a bi-exponential fit, as shown in the legend. The constant long-time value $\sigma_{n_{S},\infty}$ is the standard deviation calculated directly from the time series. The fit parameters are $a_1 = 15.08 \pm 0.01$, $a_2 = 1.81 \pm 0.01$, $\tau_1 J = 0.041 \pm 0.001$, and $\tau_2 J = 2.12 \pm 0.01$.}}    \label{fig:occupations_fluctuations_log}
\end{figure}

Thus, a direct comparison with the entropy evolution in the subsystem $(\mathcal{S}:\, i\geq 3)$ can be made only for the subsystem occupation number $n_{\mathrm{sys}}=n_3 + n_4 + n_5$. Its time evolution is shown in Figure~\ref{fig:occupations} (dashed line) along with the occupation numbers of the individual levels, $n_i$. It is seen that for short times, $Jt\simeq 1$, the occupation numbers vary strongly, but they settle to stationary values in the long-time limit. The approach of $n_{\rm sys}$ to the long-time limit, shown in Figure~\ref{fig:occupations_fluctuations_log} on a logarithmic scale, is exponential with a time scale of $\approx$10$Jt$, similar to the entanglement entropy (Figure~\ref{fig:entropy}). From Figure~\ref{fig:occupations} we also see that the long-time, individual occupation numbers systematically decrease with increasing level number $i$. We will demonstrate in the following that this indeed corresponds to a thermal distribution with a single temperature by analyzing the energy-resolved single-particle spectra.

\subsection{Long-Time Thermalization of Local Spectra}
\label{sec:thermalization_local_numbers}

In order to assess the thermalization of physical quantities, it is necessary to compute their energy-dependent spectra and to examine whether they obey a thermal distribution. Here, we are concerned with local, single-particle spectra, while energy-dependent, non-local density correlations will be analyzed in Section \ref{sec:nonlocal_correlations}.

Single-particle spectra are defined using the two-time lesser and greater Green functions (in the notation of Ref.~\cite{rammer_quantum_2011}),
\begin{align}
G_{ij}^{<}(t_1,t_2) &= -i \braket{b^\dagger_j(t_2) \, b^{\phantom{\dagger}}_{i}(t_1)}, \label{eq:Glesser}\\
G_{ij}^{>}(t_1,t_2) &= -i \braket{b^{\phantom{\dagger}}_{i}(t_1) \, b^\dagger_j(t_2)}, \label{eq:Ggreater}
\end{align}
from which the Keldysh and spectral functions, respectively, are obtained in the time domain as
\begin{align}
G_{ij}^K(t_1,t_2) &= G_{ij}^{>}(t_1,t_2) + G_{ij}^{<}(t_1,t_2)\,,\\
\,A_{ij}(t_1,t_2) &= i[G_{ij}^{>}(t_1,t_2) - G_{ij}^{<}(t_1,t_2)]\,
\end{align}

The energy-dependent Keldysh and spectral functions, $G_{ij}^K(t,E)$, $A_{ij}(t,E)$, are then obtained by transforming to center-of-motion (CoM) and relative time coordinates, \mbox{$t=(t_1+t_2)/2$,} $\tau=t_1-t_2$, respectively, and Fourier-transforming with respect to the time difference $\tau$,
\begin{align}
A_{ij}(t,E) = \int {\rm d}\tau \, A_{ij}(t_1,t_2) \, {\rm e}^{i\,E\tau}\,,
\label{eq:FT}
\end{align}
and similar for $G_{ij}^K(t,E)$. The total Keldysh and spectral functions of the system are then obtained by tracing over the level indices,
\begin{align}
G^K(t, E) = \sum_i G_{ii}(t,E), \qquad \quad
A(t,E) = \sum_i A_{ii}(t,E)\,,
\end{align}
which are independent of the choice of basis taken in level space, i.e., diagonalization is not required. In thermal equilibrium (and thus dropping the CoM time), $A(E)$ and $i\,G^K(E)$ are strictly real, positive semidefinite for $E\geq 0$, and they obey the fluctuation--dissipation relation formulated for the Keldysh function \cite{rammer_quantum_2011},
\begin{align}
G^K(E) = -i\,A(E)\,\coth\Big( \frac{E}{2 k_B T} \Big)
= -i \, A(E) \, \big( 2n_B(E,T) + 1 \big)\,.
\label{eq:fluctuation-dissipation}
\end{align}

This follows from the fluctuation--dissipation theorem for the equilibrium two-point expectation values in the frequency domain~\cite{rammer_quantum_2011} ($\beta=1/k_{\rm B}T$),
\begin{align}\label{eq:fluctuation-dissipation_general}
\braket{\hat{A}(\tau) \hat{B}(0)}_E &= e^{-\beta E} \braket{\hat{B}(0) \hat{A}(\tau)}_E\,,
\end{align}
for arbitrary operators $\hat{A}$ and $\hat{B}$ in the Heisenberg picture. The single-particle distribution function $n_B(E/T)$ can be extracted from Equation~(\ref{eq:fluctuation-dissipation}) in terms of the ratio $G^K(E)/A(E)$.

\begin{figure}[H]
\includegraphics[width=0.55\linewidth]{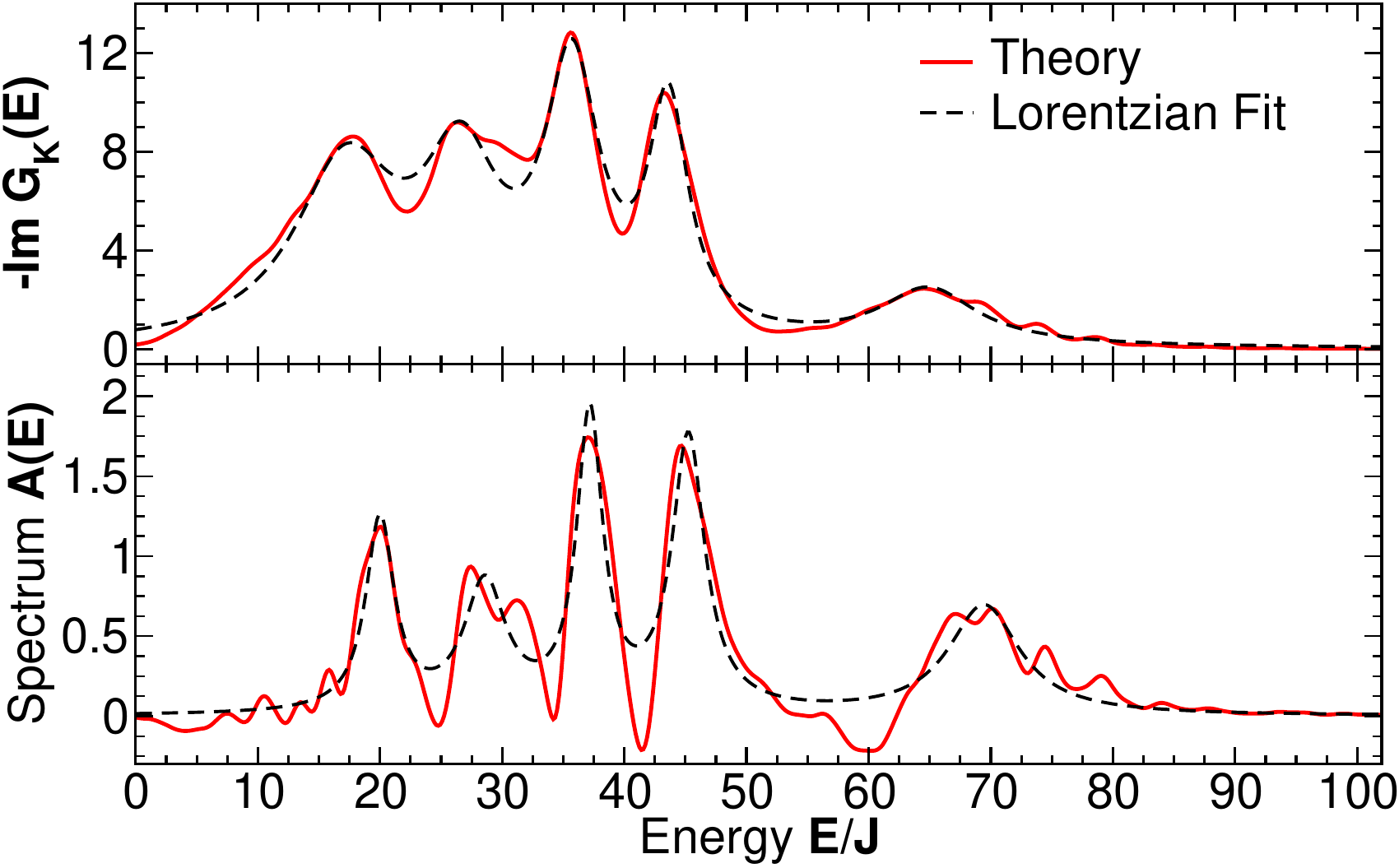}
\caption{{The spectral} function and Keldysh functions with the sum of 5 Lorentzian fits to the peaks. The center-of-motion time is $Jt = 100$.}
\label{fig:greenfunction_fits}
\end{figure}

\textls[-30]{To evaluate the two-time expectation values of Green functions in Equations~\eqref{eq:Glesser} and~\eqref{eq:Ggreater}, we write the Heisenberg operators in terms of the time-evolution operators, \mbox{$b(t_1)=\hat{U}^{-1}(t_1) \, b(0) \,\hat{U}(t_1)$,} etc., perform an appropriate, cyclic permutation under the trace to cast the time dependence on the states in the density matrix, and then evolve all the many-body state vectors appearing in the expression for $G_{ij}^{\lessgtr}(t_1,t_2)$ (before and after the action of destruction or creation operators) to the desired, long time using the algorithm described in Section \ref{sec:model_method}. A high time resolution with respect to the difference time $\tau=t_1-t_2$ is obtained by subsequently applying time-evolution operators of lower order, $\hat{U}_k$, $k < n$ (cf. Equation~(\ref{eq:t-evolution})). Finally, the trace over bath and then system states are taken for each pair $t_1$, $t_2$ with fixed CoM time $t$ and Fourier-transformed with respect to $\tau$ to obtain the spectra at time $t$.
The results for $G^K(E)$ and A(E) for a generic system of $N=25$ bosons with the broad-spectrum initial state shown in Figure~\ref{fig:spectra}, bottom panel, are shown in Figure~\ref{fig:greenfunction_fits} at a CoM time of $Jt=100$ along with the sum of Lorentzian fits to each of the five distinct spectral peaks in $i\, G^k(E)$ and $A(E)$. The slightly negative values of $A(E)$, not allowed in equilibrium, result from numerical imprecisions of the Fourier transform, Equation~(\ref{eq:FT}), and from the fact that the $\infty$-time limit cannot truly be reached numerically, so there may be slight deviations from equilibrium. Figure~\ref{fig:green_distribution} shows the spectral distribution $n_B(E,T)$ function where each data point at energy $E_i$ is extracted from Equation~(\ref{eq:fluctuation-dissipation}) as $n_B(E_i,T)= (i\,\bar{G}^K_i/\bar{A}_i)-1/2$, where $\bar{G}^K_i$, $\bar{A}_i$, $i=1,\ \dots,\, 5$, are the peak weights of the Lorentzian fit for each of the peaks in Figure~\ref{fig:greenfunction_fits} and $E_i$ are the corresponding peak positions. The solid line is a fit of the thermal Bose--Einstein distribution function $n_{\rm B}(E/T)=1/[\exp(E/T)-1]$. Note that the chemical potential $\mu=0$, since the particle number in the measured subsystem $\mathcal{S}_{i}$ is not fixed but determined by the system dynamics. Thus, the temperature $T$ is the only fit parameter.  The numerical results agree quantitatively remarkably well with a thermal distribution for all data points, with only a large error bar for level number $i=2$, resulting from the non-Lorentzian shape of this peak in the numerical result for $A(E)$. Note that the measurement (or calculation) of each level--occupation number creates a different partitioning of the total system, so that each level may, in principle, have its own temperature. However, for sufficiently long times, all these temperatures attain a global, equal value, as our single-parameter fit shows. See also CoM-time dependent results in Section \ref{sec:nonlocal_correlations}. The value of the fitted temperature is determined by the expectation value of the total energy of the initial state. We conclude that the partitioning of our generic, isolated Bose gas into observed subsystems induces a dynamical approach to the thermalization of the occupation number distribution at long times.}

\section{Thermal Behavior of Non-Local Density Correlations}
\label{sec:nonlocal_correlations}

\begin{figure}[t!]
\includegraphics[width=0.55\linewidth]{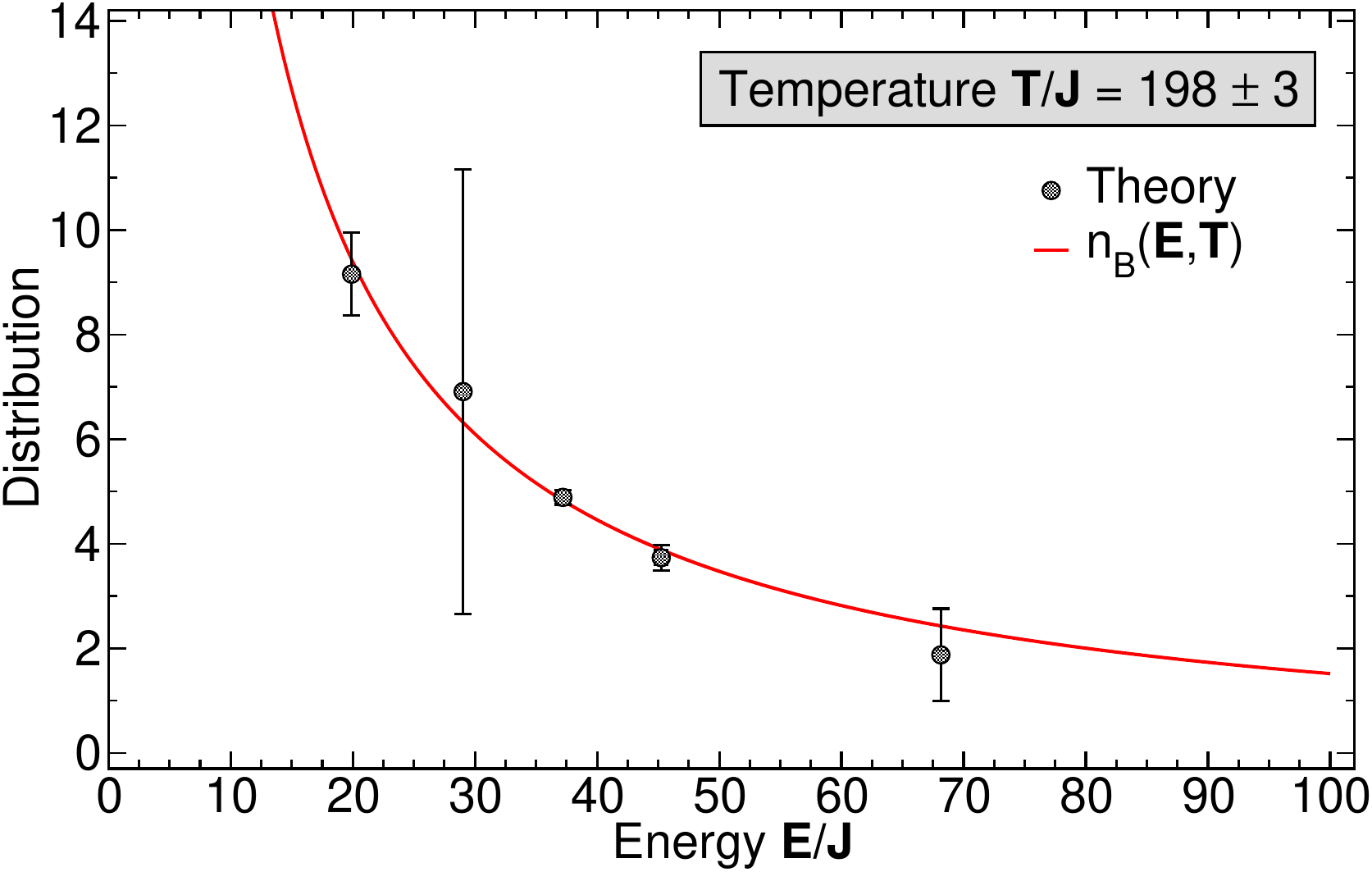}
\caption{\textls[-15]{The occupation probabilities $n_{\rm B}(E_i,T)$ of the 5 levels, $i=1,\,\dots,\ 5$ (data points), as extracted from Equation~(\ref{eq:fluctuation-dissipation}), and taking for each level the ratio of the Keldysh Green function peak weight to the spectral-function peak weight (see text). For each peak, the weight is determined as the weight of a Lorentzian fitted to the computed spectral peak. The solid curve is a Bose--Einstein distribution fitted to the data points. The only fit parameter is the dimensionless temperature $T/J=198 \pm 3$. The error bars are determined by the respective Lorentzian fits.}}
\label{fig:green_distribution}
\end{figure}

\begin{figure}[H]
\includegraphics[width=0.7\linewidth]{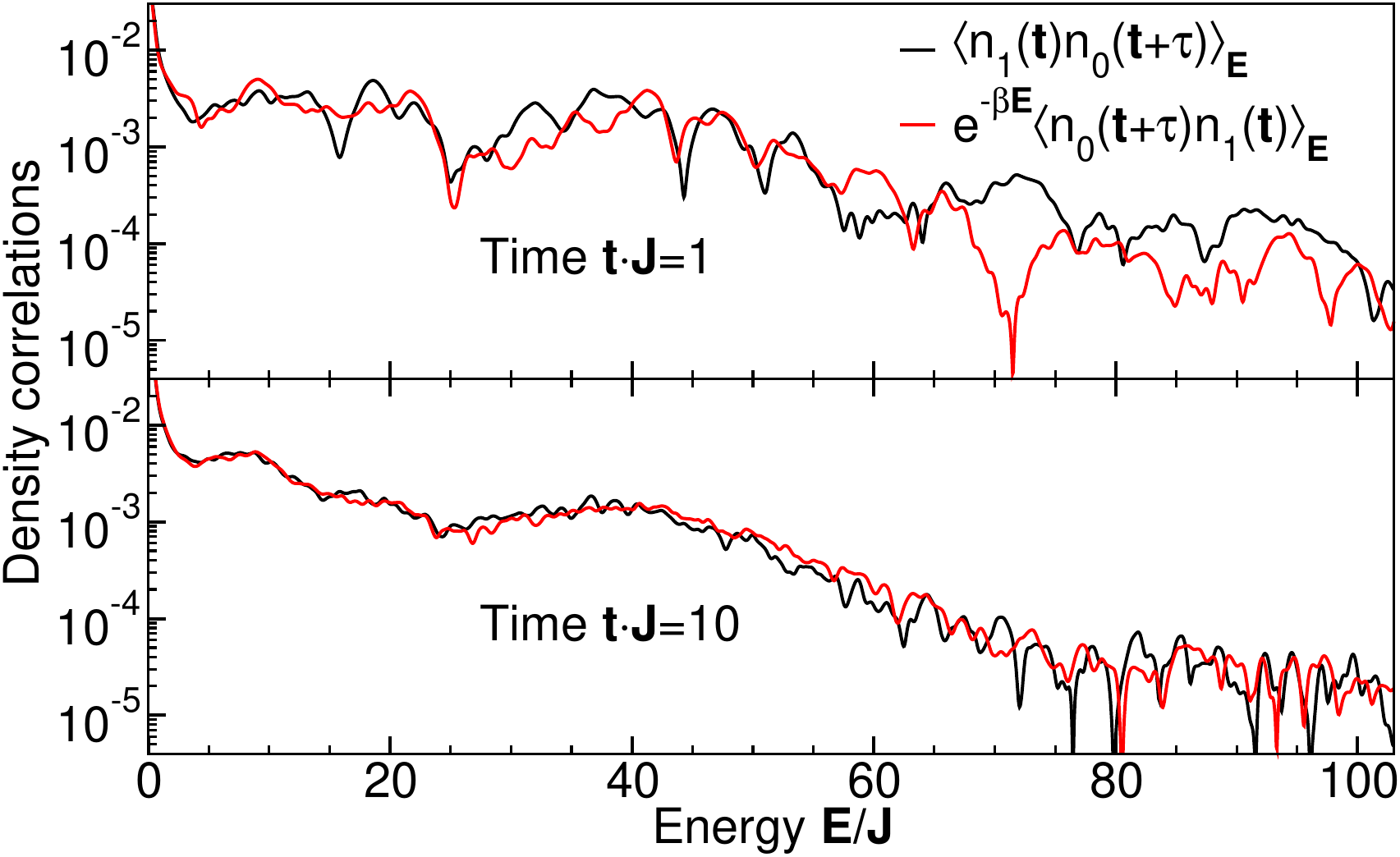}
\caption{{Fluctuation}--dissipation fits for energy-dependent density correlation functions between the lowest two levels as indicated, at times $Jt=1$ and $Jt=10$.  The inverse temperature $\beta=1/k_{\rm B}T$ is the only fit parameter. The fitted temperatures at $Jt=1$ and $Jt=10$ are $T_1/J = 193 \pm 17$ and $T_2/J = 200 \pm 7$, respectively.}
\label{fig:correl_fit_compare}
\end{figure}

Our analysis can be extended to observable quantities non-local in level space, that is, occupation-number correlation functions between different single-particle levels, \linebreak  $\chi_{ij}(t,E)=\braket{\hat{n}_i(t_1)\,\hat{n}_j(t_2)}_E$, where $\hat{n}_i(t)=b^{\dagger}(t)b(t)$ is the occupation number operator on level $i$. The thermal behavior of non-local correlation functions is usually not obtained in the ETH scenario.  Here, we calculate these quantities using the procedure described in the previous section. We probe their thermal behavior by testing them against the fluctuation--dissipation theorem in Equation~(\ref{eq:fluctuation-dissipation}). Figure~\ref{fig:correl_fit_compare} shows the comparison of the left- and right-hand sides of Equation~(\ref{eq:fluctuation-dissipation_general}) for spectra of typical occupation--density correlation functions, $\chi_{01}(t,E)$, $\chi_{10}(t,E)$, for the same parameter set as in Section \ref{sec:loc_occ_numbers}, using $\beta=1/k_{\rm B}T$ as the only fit parameter. While for short CoM evolution times $t$, the deviations are rather large at all energies $E$ (note the logarithmic scale), the agreement at longer times is very good up to $E/J\approx 60$. For even larger energies $E/J$, the density correlations decay to such small values that the deviations between both curves become comparable to the omnipresent and necessary thermal fluctuations. Hence, the non-local density correlation functions thermalize as a function of time, induced by measurement-induced partitioning. The extracted temperature of $T/J\approx 200$ of the non-local density correlation agrees well with the temperature of the local level--occupation distribution of Figure~\ref{fig:green_distribution}.

\begin{figure}[H]
\includegraphics[width=0.7\linewidth]{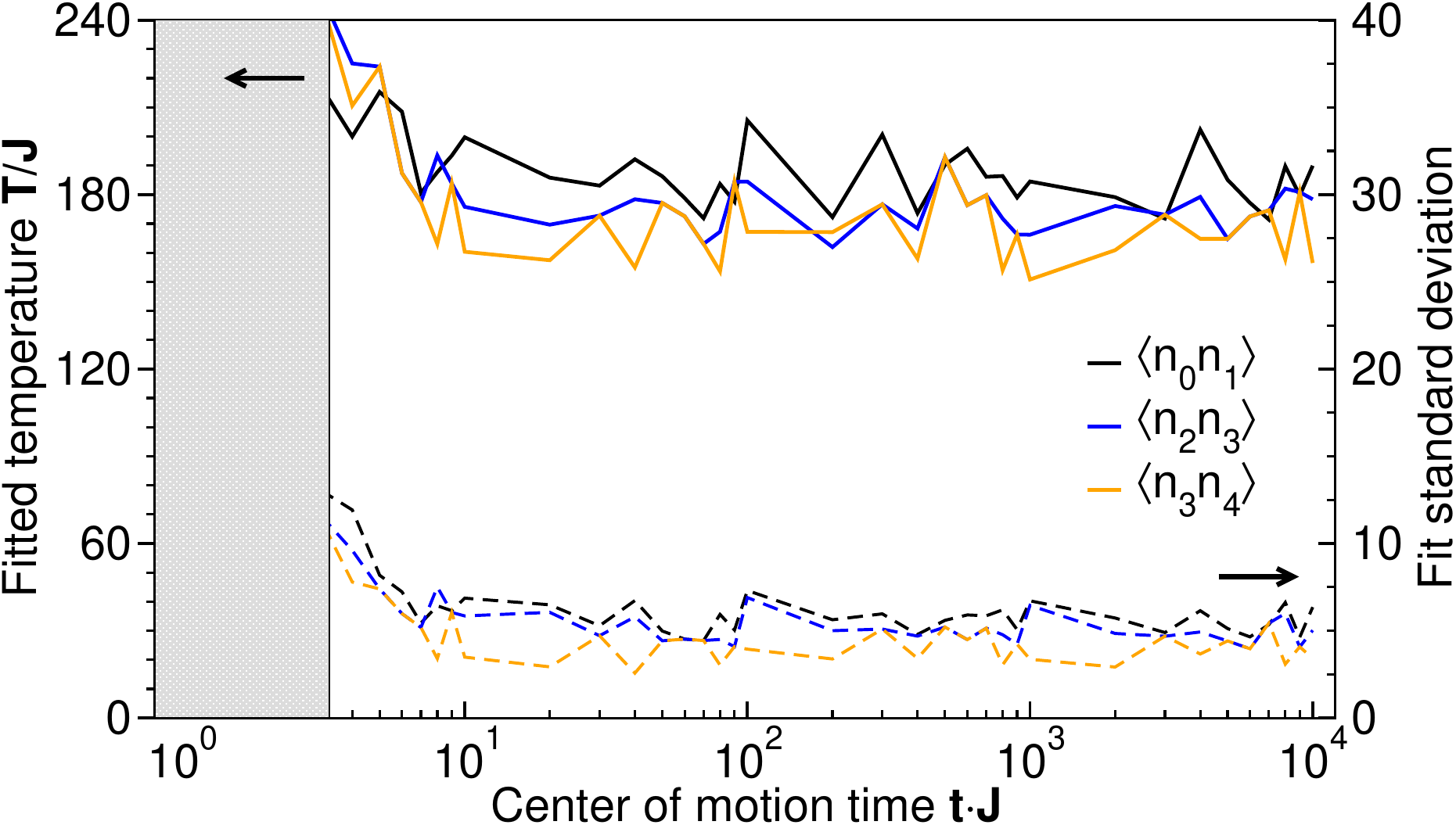}
\caption{The time‑dependent temperatures extracted from the fluctuation--dissipation fit (\mbox{Equation~(\ref{eq:fluctuation-dissipation_general}})) of level--occupation correlation functions $\chi_{i,j}$, shown as solid lines referring to the left axis. Dashed lines, referring to the right axis, represent the standard deviations of each fit. The shaded, gray area indicates the short-time region where the uncertainty of a single-temperature fit is large (see text).}
\label{fig:correl_pairs_time_dep}
\end{figure}

To exemplify the dynamical approach to thermal behavior, Figure~\ref{fig:correl_pairs_time_dep} shows the temperature as extracted from the fluctuation--dissipation fit (Equation~\eqref{eq:fluctuation-dissipation_general}) for three different correlation functions for a large range of CoM times (solid lines, left axis). For short CoM times, $Jt\lesssim 3$, a single-parameter fit of temperature is not possible, indicating that the measured subsystem is far from equilibrium. For larger CoM times, the uncertainty of the single-temperature fit becomes of the order of $3\%$ (dashed lines, right axis), indicating that a momentary temperature at time $t$ is well defined. However, at long and intermediate times, the solid lines still deviate from each other significantly more than the standard deviation of the fitting procedure (dashed lines) for different correlation functions. While this could indicate that different correlation functions may be at different, individual temperatures, the fitted mean temperature values (solid lines) fluctuate in time more than their instantaneous standard deviations (dashed lines). Within a time average over a characteristic fluctuation time, all three temperatures agree with each other, so that global thermal behavior is reached, but for exceedingly long times. The appearance of different transient temperatures will be a subject of further research.

\section{Summary and Outlook}\label{sec:conclusion}

We presented measurement-induced, dynamical bath generation (DBG) \cite{Posazhennikova2018} as a general thermalization mechanism for isolated quantum systems, which is based on ideas of canonical typicality \cite{goldstein_canonical_2006} and intrinsic decoherence \cite{zurek_decoherence_2003,milburn_intrinsic_1991,Yukalov2012_decoherence,Yukalov2012_thermalization}. In particular, we analyzed the dynamical approach to thermal behavior by numerical long-time evolution of an interacting, trapped Bose gas as a generic, non-integrable quantum system. The measurement (or calculation) of an observable quantity partitions a quantum system into observed and non-observed Hilbert subspaces in such a way that the states of the non-observed Hilbert subspace, the \textit{{bath}},
are traced over like in a grand-canonical reservoir. This tracing leads to a non-vanishing entanglement entropy of the measured subsystem with the bath despite the fact that the state vector of the entire system obeys unitary time evolution and remains a unique state for all times.

Taking a generic, non-energy eigenstate with a broad energy distribution as the initial state of the time evolution, we showed numerically that, in the long-time limit, the entanglement entropy approaches a global maximum in a bi-exponential way: The entanglement entropy reaches a near-maximum value at a rate of about one interaction constant $J$ and then evolves to the global maximum at a much slower rate. In the long-time limit, the deviations from the global maximum are dominated by thermal, statistical fluctuations. Correspondingly, all quantities considered, local observables as well as non-local correlation functions, reach thermal equilibrium distributions. Their temperatures, as extracted from the dissipation--fluctuation theorem, agree with each other within numerical accuracy. Here, locality and non-locality refer, in a more general sense, to the diagonal and off-diagonal elements of an observable not only in position, but, e.g., in level space.

When the entire system's state vector comprises an energetically narrow spectrum of many-body eigenstates, the entanglement entropy and local occupation numbers have essentially no dynamics, as expected for the microcanonical ensemble. In addition to the DBG scenario, the eigenstate thermalization hypothesis (ETH) may apply to this case in that the long-time average of the level occupation numbers is equal to the microcanonical ensemble average for an ergodic system \cite{deutsch_quantum_1991,srednicki_chaos_1994}.

These results allow for a number of general conclusions.
The DBG scenario appears to have no conceptual restrictions on its applicability: In contrast to the ETH, it describes the time-dependent approach to thermal behavior. This is true even for an initial state with the lowest possible entropy, a pure state, and works for broad as well as for narrow energy distributions in the initial state. Furthermore, we showed that the DBG is not limited to the thermalization of local (in level space) but also applies to non-local correlation functions. The only condition on the applicability appears to be that the system is sufficiently complex and not integrable, that is, the Hilbert space has sufficiently large dimension and does not factorize into disconnected sectors by Hamiltonian dynamics. In particular, we conjecture that it is, in fact, the large Hilbert space dimension that ensures a well-defined thermodynamic limit, rather than the particle number. This is because our calculated expectation values are well defined in that their temporal variations are much smaller than the mean value in the long-time, stationary limit, even though the particle numbers $N\leq 25$ of our system are not macroscopic. The bi-exponential approach of the entanglement entropy to a global maximum found in our calculations suggests an intermediate, near-equilibrium state. It will be a subject of further research in how far this may be described by time-dependent temperatures of subsystems at intermediate times.

\vspace{6pt}
%
%
%
\authorcontributions{\textls[-15]{Conceptualization, A.P. and J.K.; methodology, M.L., S.B. and J.K.; software, M.L. and S.B.; validation, M.L. and J.K.; formal analysis, M.L. and S.B.; investigation, M.L. and S.B.; resources, J.K.; data curation, M.L.; writing---original draft preparation, M.L. and J.K.;} writing---review and editing, M.L., S.B., A.P. and J.K.; visualization, M.L.; supervision, A.P. and J.K.; project administration, J.K.; funding acquisition, J.K. All authors have read and agreed to the published version of the manuscript.}

\funding{This work was supported by the Deutsche Forschungsgemeinschaft (DFG, German Research Foundation) under Germany’s Excellence Strategy---Cluster of Excellence “Matter and Light for Quantum Computing”, ML4Q (390534769), and through the DFG Collaborative Research Center CRC 185 OSCAR (277625399). S.B. acknowledges financial support by the Deutsche Akademische Austauschdienst (DAAD, German Academic Exchange Service) through the Working Internships in Science and Engineering (WISE) Program, 2019 (57460839).}

\dataavailability{{The data} presented in this study are openly available in Zeneodo at \url{https://zenodo.org/records/15655794}  (accessed on 24 May 2025).
}

\acknowledgments{\textls[-15]{We are grateful to Tim Bode, Sayak Ray, and Michael Turaev for useful discussions.}}

\conflictsofinterest{The authors declare no conflicts of interest.}


\abbreviations{Abbreviations}{
The following abbreviations are used in this manuscript:\vspace{3pt}

\noindent
\begin{tabular}{@{}ll}
ETH & Eigenstate thermalization hypothesis\\
DBG & Dynamical (heat) bath generation\\
FDT & Fluctuation--dissipation theorem\\
GOE & Gaussian orthogonal ensemble\\
COM & Center of motion
\end{tabular}
}
\begin{adjustwidth}{-\extralength}{0cm}
\reftitle{{References} 
}

\PublishersNote{}

\end{adjustwidth}
\end{document}